# An Approach to Implement Photovoltaic Self-Consumption and Ramp-Rate Control Algorithm with a Vanadium Redox Flow Battery Day-to-Day Forecast Charging

Ana Foles[a,b,1], Luís Fialho[a,b,2], Manuel Collares-Pereira[a,b,3], Pedro Horta[a,b,4]

[a]*Renewable Energies Chair, University of Évora. Pólo da Mitra da Universidade de Évora, Edifício Ário Lobo de Azevedo, 7000-083 Nossa Senhora da Tourega, Portugal*
[b]*Institute of Earth Sciences, University of Évora, Rua Romão Ramalho, 7000-671, Évora, Portugal*
[1]anafoles@uevora.pt
[2]lafialho@uevora.pt
[3]collarespereira@uevora.pt
[4]phorta@uevora.pt

Nomenclature

*Abbreviation, Definition*

| | |
|---|---|
| API | Application Programming Interface |
| AROME | Application of Research to Operations at MesoscalE |
| BAPV | Building Applied Photovoltaics |
| BCR | Battery Charge Ratio |
| BESS | Battery Energy Storage System |
| BTN | Normal Low-Voltage |
| CRR | Controlled Ramps Ratio |
| DL | Decrew-Law |
| DOD | Depth of Discharge (%) |
| DSM | Demand Side Management |
| DSO | Distributor System Operator |
| ECMWF | European Centre for Medium-Range Weather Forecasts |
| EDP | Electricity of Portugal |
| EG | Energy from the Grid |
| EMS | Energy Management Strategy |
| ENTSO-E | European Network of Transmission System Operators for Electricity |
| ESS | Energy Storage System |
| FBU | Overall From Battery Use |
| FGU | Overall From Grid Use |
| GRF | Grid Relief Factor |
| IPMA | Portuguese Institute for Sea and Atmosphere |
| JSON | JavaScript Object Notation |
| KPI | Key-performance Indicator |
| MA | Moving Average |
| MPP | Maximum Power Point |
| MPPT | Maximum Power Point Tracker |
| PV | Solar Photovoltaic |
| RFB | Redox Flow Battery |
| RR | Ramp Rate (%/min per nameplate capacity) |
| SCM | Self-Consumption Maximization |
| SCM+RR | Self-Consumption Maximization with Ramp Rate Control |
| SCM+RR+WF | Self-Consumption Maximization with Ramp Rate Control with Weather Forecast VRFB Charging |
| SCR | Self-Consumption Ratio |
| SOC | State of Charge (%) |
| SSR | Self-Sufficiency Ratio |
| TBU | Overall To Battery Use |
| TGU | Overall To Grid Use |
| TMY | Typical Metereological Year |
| TSO | Transmission System Operator |
| UÉvora | University of Évora |
| UPS | Uninterruptible Power Supply |
| UTC | Coordinated Universal Time |
| VRE | Variable Renewable Energy |
| VRFB | Vanadium Redox Flow Battery |
| WF | Weather Forecast |

Abstract

The variability of the solar resource is mainly caused by cloud passing, causing rapid power fluctuations on the output of photovoltaic (PV) systems. The fluctuations can negatively impact the electric grid, and smoothing techniques can be used as attempts to correct it. However, the integration of a PV+VRFB to deal with the extreme power ramps at a building scale is underexplored in the literature, as well as its effectiveness in combination with other energy management strategies (EMSs). This work is focused on using a VRFB to control the power output of the PV installation, maintaining the ramp rate within a non-violation limit and within a battery state of charge (SOC) range, appropriate to perform the ramp rate management. Based on the model simulation, energy key-performance indicators (KPI) are studied, and validation in real-time is carried. Three EMSs are simulated: a self-consumption maximization (SCM), and SCM with ramp rate control (SCM+RR), and this last strategy includes a night battery charging based on a day ahead weather forecast (SCM+RR+WF). Results show a battery SOC management control is essential to apply these EMSs on VRFB, and the online weather forecast proves to be efficient in real-time application. SCM+RR+WF is a robust approach to manage PV+VRFB systems in wintertime (studied application), and high PV penetration building areas make it a feasible approach. Over the studied week, the strategy successfully controlled 100% of the violating power ramps, also obtaining a self-consumption ratio (SCR) of 59% and a grid-relief factor (GRF) of 61%.

1. Introduction

At the end of 2019, the global installed renewable energy capacity reached 2,537 GW, more 176 GW compared with 2018 [1]. In the same year, solar photovoltaic (PV) energy had a 3% share in the world generation mix, with a 2050 forecast of 23% [2]. With the significant share of renewable energy, which is variable (VRE), it is necessary to impose limits for its integration into the grid [3]. The integration of VRE obeys a regulation that defines the conditions of the parameters to be exchanged with the grid (quality). The parameters are, for instance, the voltage and frequency operating ranges, reactive power capacity for voltage control, active power gradient limitations, among others. The fluctuation of primary energy from VRE is concerned with the limitations of the active energy gradient. The increase in VRE's contribution to final energy consumption can be resolved by creating a ramp rate limit, which is already in the legislation of some countries around the world [3]. Countries with a significant number of solar PV installations have limitations in supplying energy to the grid. The ENTSO-E (European Network of Transmission System Operators) requires the ramp rate to be specified by the regional Transmission System Operator (TSO), if necessary. In Germany, the grid code establishes that if the installed capacity of a PV generator is greater than 1 MVA, the ramp rate limit is 10% of the rated power per minute. In Puerto Rico island, the PREPA 2012 regulation imposes as well a 10% of nameplate capacity per minute as a ramp rate limitation of grid injection [4]. In Ireland, the EirGrid Plc regulation establishes that the wind farm power stations must be able to control the ramp rate of their active power output over a range of 1-100% of their nominal capacity per minute. The wind turbines must be able to restrict ramping [3]. In the Philippines, the grid code establishes an active power limit during over-frequency. The largest plants must be able to limit ramps, and in China, a National Standard was created to control the maximum ramp range of PV power stations to be less than 10% of the installed capacity per minute [5].

Short-term energy fluctuations are directly related to the area of the PV plant and its geographical dispersion, and for this reason, in general, the small area, associated with a building PV installation, becomes especially subject to severe fluctuations in PV energy [6]. Ramp rate negative impact is related to matters as the extension of household appliances lifetime, and the contribution to the grid stability. To characterize the impact that the 6740 W of one of the UÉvora Building Applied PV (BAPV) systems has on the building, a calculation was made for the ramp rate, for 1 minute with one year of PV recorded data (2018). The monitoring data is collected through an in-house developed software using the LabVIEW environment, with a 2-second time frame, using a precision power analyser (Circutor CVM-1D [7]) and the PV inverter. The ramp rate was calculated for the entire year for values of ramp rates of 5% and 10% as current references, and the results are shown in Table 1, below.

Table 1 – Ramp rate in %/min of the nameplate capacity of the PV system (UÉvora, 2018 data).

| Ramp rate (%/min) in the year 2018 | Percentage of total ramp rate in one year (%) |
|---|---|
| < 5 %/min | 78.5 |
| ≥ 5 %/min | 8.09 |
| ≥10 %/min | 5.16 |
| > 10 %/min | 4.83 |
| ≥ 50 %/min | 0.65 |

From the observation of Table 1, about 8% are above the 5%/min ramp rate and about 5% are higher or equal to 10%/min of the PV nameplate capacity, which is an expected result given the PV installation size. These ramp rate values have probably low impact on the grid, although should not be ignored. To increase the degree of confidence of these results, a substantially greater data period is required, ideally, several years. The existence of systems with monitoring of long-term PV generation data with high frequency (as, for instance, data logging at 2s) is rare and should be an effort to be implemented in the future of experimental installations. The results of Table 1 are representative of the location and specifications (tilt, azimuth, among others) of this single PV system, due to the direct relationship between this solar radiation data and the occurrence of power ramps, as the analysis of the solar radiation meteorological data.

To tackle the PV fluctuations the ramp rate control can be achieved through three main techniques, namely, operation of the PV system below its nominal capacity, bypassing the MPP [8], or through the use of a Battery Energy Storage System (BESS) to absorb or inject the excess of generated PV energy when the ramp rate is violated [9]. BESS helps in the regard of self-consumption and self-generation of energy and grid auxiliary. It also contributes to energy loss reduction, reliability increase, stability, power quality increase and energy efficiency, help in the systems operation and frequency regulation and balance establishment among energy demand and supply [10]. Ramp rate limitations are generally studied to apply in large PV installations, where their effects are more noticeable, due to the reduction or sudden increase in the power injected into the grid. Although, with the increase of PV installations number in the buildings sector, this issue should be addressed. In the literature, the authors of [11] approach the ramp rate control application to deal with PV fluctuations at a real scale using a BESS to compensate sizing to deal with ramping. The authors of [9] explore the PV ramping application with a battery state of charge (SOC) reference calculation. Previous works are references in the field, although there is a clear lack of studies devoted to the domestic and services (buildings) sector and real-time application, given the previous studies focused on large-scale solar PV plants and BESS sizing. The authors of [12] developed the ramp rate and self-consumption algorithm validation with a DC controller for an ultracapacitor as BESS in a microgrid, although the building integration at real scale is still lacking and being the control devoted for DC microgrids. The authors of [13] present the comparison of different ramping smoothing filter techniques using a BESS, exploring time intervals of 2.5, 5, 7.5, and 10 minutes, evaluating the number of fluctuations and their impact on the application, and the authors of [14] consider a time interval of 15 minutes. Both works have considered large sets which is not the case of the present work, where the time frame of seconds is considered, as further discussed. On the contrary, the research conducted by [15] considers a 250 ms PV and wind data, explaining the lower intervals impact.

A BESS is generally managed optimally if applying energy management strategies (EMSs) in its control and operation, involving the generation and the consumption of energy. PV for self-consumption is the most studied and applied strategy to operate a PV system with or without a battery (depending on the consumption needs). In the context of the Portuguese legislation in force, the Decree-Law 169/2019 [16], PV self-consumption is highly promoted, being a starting point of this work. Within this topic, a greater volume of works has been developed, not only for PV-only configurations, as much as integrating a BESS, at buildings scale, with different final objectives improvements, namely, economic, technic, or energetic. The variety of works in literature present energy management strategies as the PV self-consumption maximization in [17], demand-side management (DSM) in [18] and [19], the use of load forecasts to optimize the operation of PV and the battery in [20], load-scheduling [21], peak-shaving and power curtailment [22], or battery charging controls [23] as battery operating scheduling [24], among others. In the case of the present work, the more significant EMS under study is the self-consumption maximization combined with a ramp rate control, considering a PV installation and a BESS.

In this work, the BESS technology under study is a redox flow battery (RFB), considered promising for stationary energy storage in electric grids [25]. The RFB electrochemical processes occur as redox reactions in its conversion unit, the stack. The stack is made of several cells, which form two electrodes separated by a proton selective membrane. The electrodes are made of porous graphite felts, and the bipolar plate between each cell creates the electrical connection between the two opposite poles. The electrolyte is pumped from the tanks to the stack, where the half-electrochemical reactions occur. RFB technology has advantages in the decoupling of power and energy ratios, large cycle life, low maintenance, and limited self-discharge. As for disadvantages, it has a low power density. Different chemistries of RFBs exist, as the example of zinc-bromine, hydrogen-bromine, among others, although vanadium-vanadium chemistry is the most mature so far, introduced in the 1970s and already commercialized.

The vanadium redox flow battery (VRFB) has the vanadium element in four oxidation states mixed in an aqueous solution of sulfuric acid. The storage of energy is made in two electrolytic solutions with two different redox couples: the negative electrode is composed of bivalent $V^{2+}$ and trivalent $V^{3+}$ ions; the positive electrode is composed of tetravalent $VO^{2+}$ and pentavalent $VO_2^+$ ions [26]. VRFBs are suited for applications requiring security of energy supply, energy/power quality, load levelling, and renewable energy compensation as time-shift, grid efficiency and off-grid applications [27], peak shaving, and uninterrupted power supply (UPS). Details of costs can be consulted in [28] – a 1 MW/4 MWh VRFB ESS states at 391 $/kWh in the 2020 year, with a projection of 318 $/kWh for the 2030 year. In literature, VRFB as a BESS is being investigated, and in the following, the most relevant for this context are highlighted. In the scale of MW VRFB, the potential integration of VRFBs with wind farms is investigated by the authors of [29] to mitigate power grid and market integration issues, and although the simulation results of the economic study are compelling, the solution is not validated at a real-scale. In [30] the authors combine a 1 kW/6 kWh VRFB with a 10 kW solar PV, a 15 kVA biomass infrastructure, and a 1kW of wind for different EMSs purposes, and operation control is designed and simulated through LabVIEW, although the study lacks general EMSs evaluation indicators. The same issue is lacking in the work developed by the authors of [31], with relevant inputs of controlling the charging current and flow rate of the 6 kWh/ 1kW VRFB integrated with a PV maximum power point tracker (MPPT). The work is validated through some hours of operation. Authors of [32] explore modelling and operation of a VRFB strictly for PV application. There, a PV+VRFB model is explored, based on its typical operation characteristics, in conjunction with a charge controller unit, to ensure safe operation. The simulation is carried to the battery balancing the load to ensure firm power output at the load, although the real-time validation and building integration is missing.

In this work, three energy management algorithm strategies are studied, explained in the following. Self-consumption maximization (SCM – strategy 1) is a strategy that maximizes the usage of the PV generation throughout the day and can benefit from the VRFB to increase the PV power consumption. The second strategy is self-consumption maximization with ramp-rate control (SCM+RR – strategy 2) is strategy 1 including a ramp rate control accommodated by the VRFB if it presents suitable capacity (within the allowed SOC range) and power, and any SOC control is carried. The third strategy is a Self-consumption maximization with Ramp-rate control and VRFB charging based on a weather forecast (SCM+RR+WF – strategy 3) and is strategy 2 with a battery SOC control implementation. This work aims to control PV fluctuation and still execute self-consumption to benefit from the Portuguese legislation in force. The strategy is a theoretical simple approach, and if the VRFB SOC control turns effective with the local weather forecast (WF), the VRFB application is effective as well and can be applied to any VRFB+PV self-consumption scenario. Using UÉvora's microgrid as a building sector case study, this work investigates the possibility to develop and implement the real-time multipurpose algorithm control, to improve the integrated system, assuring the security of supply and solving a predicted regionally high PV penetration scenario already in force in some countries, in this case for the location of Évora, Portugal. The evaluation is carried through dedicated key-performance indicators (KPI), to check if the WF adding on the PV+VRFB operation improves (not worsens) any of the studied KPI. In this work, the exploitation of the conjunction of the SCM and the ramp rate control with the local daily WF with a VRFB as a solution for building integration is a novelty, both in simulation, as well as in the application.

This work is structured as follows: Section 2 presents the overall methodology of the work, starting with the description of the microgrid architecture and setup (2.1), as well as the VRFB modelling (2.2). Section 2.3 presents the PV generation and load profiles used for the EMSs simulations. The next section (2.4) shows the ramp rate calculation method and the filter technique used and the time frame chosen for the ramp rate control effectiveness. To assess the EMSs' performance, KPIs were defined and used, as depicted in Section 2.5. Section 2.6 presents the original control implementation methodology, divided into three main sections: the upfront VRFB WF online explanation for implementation (2.6.1), the VRFB operational constraints for LabVIEW implementation (2.6.2) and the algorithm implementation section which better details its functionality (2.6.3). Section 3 provides the main simulation results of the work and its related discussion, as the resulting of the SOC control (3.1), the energy KPIs (3.2), and other related outcomes from the experimental approach (3.3 and 3.4), with a final remark on further work (3.5). Finally, the fundamental conclusions of this work are detailed in Section 4.

## 2. Methodology

The work approach follows the steps of Figure 1, which specifies the input parameters subjacent to the simulation model and the output of the energy analysis (A), and the details regarding the experimental validation for the EMS's feasibility (B).

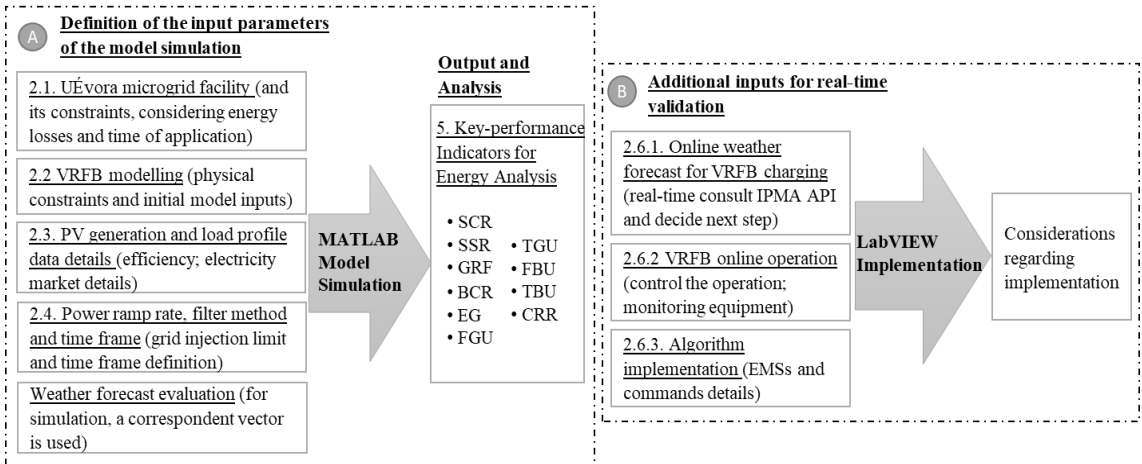

Figure 1 – Overview of the architecture underlying present work methodology. The definition of KPIs is detailed in Section 2.5.

The simulation of the three EMSs algorithms is developed in MATLAB. The inputs of the simulation model include the constraints details of the UÉvora microgrid, the PV generation and load profiles, the developed VRFB electrical model, and the ramp rate calculus and its premises. The evaluation of each strategy is made through dedicated energy KPIs. After the evaluation and optimization of the simulation model, SCM+RR+WF – strategy 3 is implemented in real-time using LabVIEW in the UÉvora microgrid, where the required inputs of the weather forecast are online consulted. Hence, the validation of the proposed EMS is achieved, through control with the monitoring equipment. Each of the marked text points of Figure 1, concerning the MATLAB simulation modelling (A) and the LabVIEW implementation (B), that allowed the construction of the modelling and the real-time microgrid operation are presented in the subsequent sections: the UÉvora microgrid facility corresponds to subsection 2.1, and so on successively.

## 2.1. UÉvora microgrid facility

The building scale VRFB (5 kW/ 60kWh) from redT is integrated into a three-phasic microgrid test facility of Renewable Energies Chair of the University of Évora, in operation since 2012. The roof facility is equipped with a polycrystalline technology of 3.5 kWp and a monocrystalline technology of 3.2 kWp PV systems, separated by strings, directed through another AC/DC Ingeteam PV inverter. AC metering is installed in several points of the microgrid (before and after each piece of equipment), and the DC measurements of voltages, currents, and temperature are achieved through a data precision multimeter. The power management system that operates the VRFB is composed of bidirectional inverters from Ingeteam, with an operation above 48 DC voltage, which also executes AC and DC data measurement. A desktop computer equipped with the LabVIEW environment is the control unit, which in real-time operation, gather the data under the chosen (or possible) time frame. Its main components are schematized in Figure 2, below.

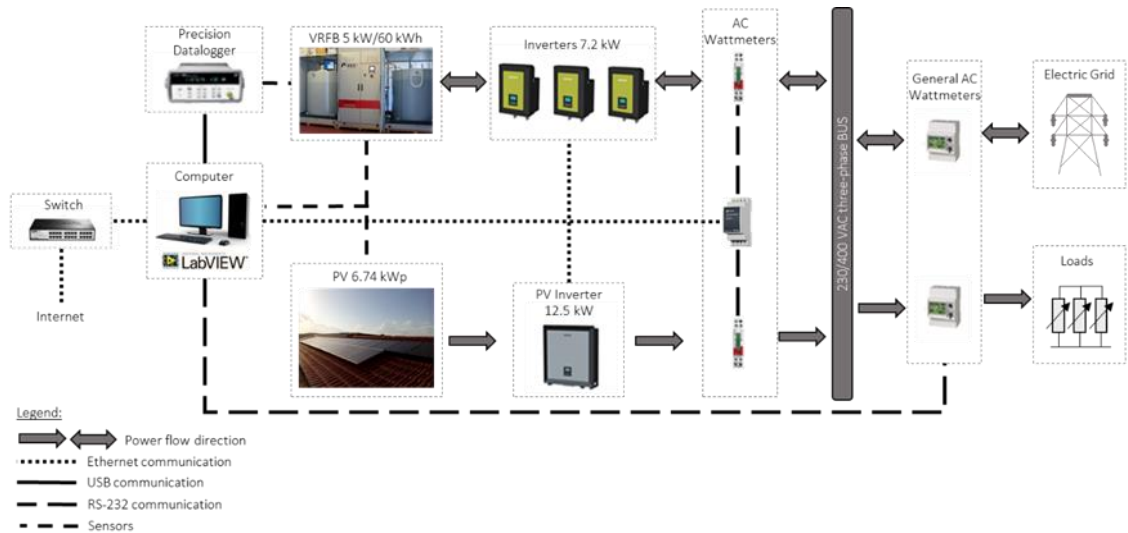

Figure 2 – Representation of UÉvora's VRFB microgrid equipment's connection and energy fluxes.

The microgrid general equipment's measurement uncertainty and communication delays should be accounted as a constraint of the real-time application. The AC/DC efficiency of the VRFB's integrated inverters was included in the developed simulation model (AC/DC efficiency of 0.88, resulting from previous experiments). The standby value of 30 W was considered.

2.2. VRFB modelling

For the simulation of the EMSs through MATLAB, a dedicated VRFB model was included. VRFB architecture allows its replication in other facilities, with the increased advantage of being an easily scalable product. The technology has two electrolyte tanks containing a mixture of vanadium ions and sulfuric acid, two pumps for electrolyte flowing, and the stack as the energy conversion unit, with 40 cells electrically connected in series, and hydraulically connected in parallel. The 5-kW stack is a dynamic system, and its performance depends on multiple effects: electrochemical, fluid dynamics, electric and thermal. This specific 60 kWh VRFB was the object of study in previous works, and the most recent include the battery electrical modelling, developed, and validated on a real operation scale, considering its general operating conditions in the UÉvora. The model is fully detailed in the research conducted in [33], and it was implemented and used in the developed simulation model to test the implementation of these EMSs. The use of this validated battery model helps the control implementation with the smallest possible error.

2.3. PV-generation and load profile data

To compose the buildings scenario, the PV profile used corresponds to the data obtained with the UÉvora's PV installations, during one week from the 1st to the 7th day of January of 2018, with 2-second intervals data logging. The load profile used is made available by EDP Distribuição – Portuguese DSO company – with 15 minutes of average load data for the year 2018 [34]. With the help of the MATLAB software, the data were treated to correspond to the PV data sampled time frame. The PV and load profiles collected data for the first seven days of January 2018 can be seen in Figure 3, below.

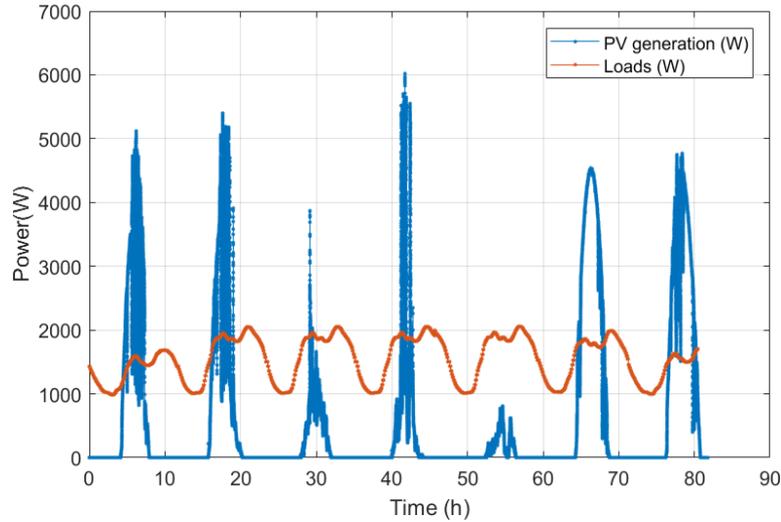

Figure 3 – Solar PV profile obtained by the measured data of the UÉvora's PV installation, and load profile, obtained by [34] estimation for BTN B sector, for 1 week of January, from day 1-7 day of 2018.

## 2.4. Power ramp rate, filter method and time frame

In general, the classic way of representing the ramp rate is defined as RR, as presented in Eq. (1),

$$RR = \frac{\frac{P_{PV}(t) - P_{PV}(t - \Delta t_R)}{P_N}}{\Delta t_R} \times 100 \quad (1)$$

where, $P_{PV}$ – PV power (W) and $t - \Delta t_R$ – time differential of the ramp rate, typically equal to the unit (min).

In this work, the authors focus on the Moving Average (MA) filter technique type, which is considered the most traditional filter technique, although it can lead to increased battery cycling. This is also the reason why the present study was chosen to apply the MA filter to a VRFB, with a high energy capacity of 60 kWh. This BESS technology, although a battery with moderate energy density, allows (if necessary) a high number of cycles without significantly reducing its performance, capacity, or life. The moving average time frame should be sensibly weighted. The contribution developed in [35] investigates the PV time averaging impacts on the small and medium-sized PV installations and concludes that a time frame of 15-minute averages describes the ramps poorly. For the week considered in this study, the theoretically controlled ramps were calculated using the 10%/min ramp rate with an average of different periods of PV intervals. The obtained values are presented in Figure 4, below.

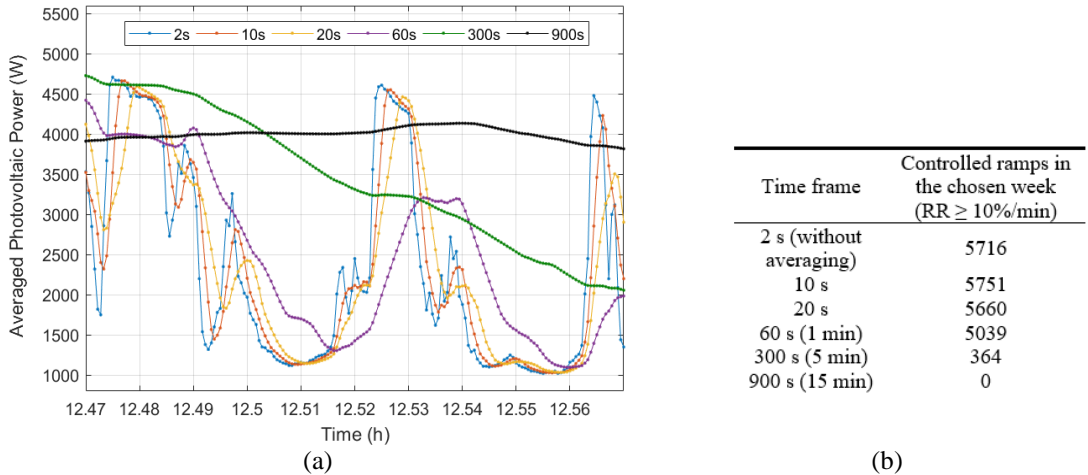

| Time frame | Controlled ramps in the chosen week (RR ≥ 10%/min) |
| --- | --- |
| 2 s (without averaging) | 5716 |
| 10 s | 5751 |
| 20 s | 5660 |
| 60 s (1 min) | 5039 |
| 300 s (5 min) | 364 |
| 900 s (15 min) | 0 |

(a)          (b)

Figure 4 – (a) PV average of the time frames studied. The 2s correspond to the raw data extracted from the PV installation under study. The remaining time frames are the PV averaging corresponding to each of that time frames. (b) The number of controlled ramps for the studied week, over different time frames of PV average values.

Figure 4 (b) results obtained support the study previously referred to in [35], for the domestic PV installations: as the average time frame increases, fewer ramps are detected and controlled. Alternatively, if the average time frame is too small, the impact will be low. The simple implementation of MA and low computational effort can let through unexpected artefacts such as peaks in the results. In accordance, and considering the microgrid constraints, a time frame of 20 seconds was chosen. This method functions with the averaging of the previous PV measurements, in a chosen period, $t$. The battery command, $P_{battery}$, at a $k^{th}$ instant, can be calculated from Eq. (2) [13]:

$$P_{battery}(k) = \frac{\sum_{i=0}^{t-1} P_{PV}(k) - P_{PV}(k-i)}{t} - P_{PV}(k) \qquad (2)$$

### 2.5. Energy Key-Performance Indicators

To properly evaluate the EMSs, suited key-performance indicators are calculated. The parameters are based on the sum of the energy used throughout the days of the strategy application. In the following Table 2, the indicators are enunciated, below.

Table 2 – Energy key-performance indicators for the evaluation of the EMS.

| KPI | Description | Equation |
| --- | --- | --- |
| Self-consumption ratio (SCR) | Share of the PV energy consumed within the installation over the total PV energy generated. The PV energy produced can be consumed indirectly by the battery (including losses). | $SCR = \frac{E_{PVconsumed}}{E_{PVgenerated}}$ |
| Self-sufficiency ratio (SSR) | Share of the consumed PV energy generation over the total consumption needs. | $SSR = \frac{E_{PVconsumed}}{E_{Load}} = \frac{E_{Load} - E_{Grid}}{E_{Load}}$ |
| Grid-relief factor (GRF) | The measure of the total energy of the installation loads, which is not exchanged with the grid. | $GRF = \frac{E_{Grid}}{E_{Load}}$ |
| Battery charge ratio (BCR) | Total energy used to charge the battery, over the overall energy sent and received to/by the battery. | $BCR = \frac{E_{charge}}{E_{battery}}$ |
| Energy from the grid (EG) | Amount of energy extracted from the grid, considering the total energy exchanged with the grid | $EG = \frac{E_{fromGrid}}{E_{Grid}}$ |
| Overall from grid use (FGU) | Amount of energy extracted from the grid in the overall installation consumption needs | $FGU = \frac{E_{fromGrid}}{E_{Load}}$ |
| Overall to grid use (TGU) | The amount of energy injected into the grid, over the overall installation consumption needs | $TGU = \frac{E_{toGrid}}{E_{Load}}$ |
| Overall from battery use (FBU) | Amount of energy extracted from the battery in the overall installation load profile | $FBU = \frac{E_{fromBattery}}{E_{Load}}$ |
| Overall to battery use (TBU) | Amount of energy sent to the battery, over the overall installation load profile | $TBU = \frac{E_{toBattery}}{E_{Load}}$ |
| Controlled ramps ratio (CRR) | Rate of the total number of the ramps (up and downs) controlled with the use of the battery using the EMS, over the total number of ramps (up and downs) without the use of an EMS | $CRR = \frac{Nr_{strategy}}{Nr_{original}}$ |

Where: $E_{PVconsumed}$ is the PV energy generation consumed directly or indirectly; $E_{PVgenerated}$ is the total PV energy generated by the PV installation; $E_{Load}$ is the total load consumption; $E_{Grid}$ is the injected and extracted energy to/from the grid; $E_{battery}$ represents the total energy sent to charge and discharge the battery, in absolute values; $E_{fromGrid}$ is the energy needed to extract from the grid to supply the energy consumption needs, in the overall strategy; $E_{Grid}$ the total amount of energy exchanged with the grid; $E_{toGrid}$ is the sum of the energy sent to the grid; $E_{fromBattery}$ is the energy used to discharge the battery; $E_{toBattery}$ is the energy used to charge the battery; $Nr_{strategy}$ is the total number of ramps controlled using the EMS; and $Nr_{original}$ the number of ramps that occur without an EMS.

## 2.6. Control Implementation

The proposed strategy 3 combined with the online WF and the VRFB is a complex task to orchestrate in real-time operation. Through the seven-day experimental validation, the feasibility of its application in real-time can be properly assessed. PV fluctuation has a seasonal-dependent characteristic, occurring in Portugal mostly in periods from winter to spring. By these means, the authors decided to evaluate this strategy application in one particularly fluctuation week (1-7[th] of January). The real-time operation of the SCM+RR+WF (strategy 3) and experimental validation details are presented in this section. The study made in [36] validated the SCM (strategy 1) using the same experimental setup and battery. Strategy number 2 is a simplification of strategy number 3. Being the SCM+RR+WF (strategy 3) the most complex strategy, both on algorithm or control implementation, when compared to strategy 1 and strategy 2, it was decided to implement the SCM+RR+WF only and validate it at full scale and real operating conditions in the experimental microgrid of the VRFB, using LabVIEW, a graphical programming code, which combines visualization results and interactive tools to allow a real-time controller interface, check the course of the algorithm and act if needed.

### 2.6.1. Weather forecast for VRFB charging

In strategy 3, SCM+RR+WF, the battery will be charged to values near the 50 % SOC when needed, with data input from weather forecast (ramp-rate occurrence), using data forecast produced by IPMA (*Instituto Português do Mar e da Atmosfera*). IPMA is a Portuguese public body, which is responsible for, among other many tasks, forecasting the states of the weather and sea, for all necessary needs. The forecasted data, associated with the geographical and seismic events, are made available in their Application Programming Interface (API) in a JSON format [37]. The data is obtained automatically through a forecast statistic process with forecasts of two numerical models – ECMWF (European Centre for Medium-Range Weather Forecasts) [38] and AROME (Application of Research to Operations at MesoscalE) [39]. These forecasts are updated two times per day, at 00 UTC (available at 10 am) and 12 UTC (available at 8 pm). In the summertime, the Portuguese legal hour is UTC+1, and in the wintertime, the legal hour is equal to UTC. In the referred online API, daily meteorological data forecast up to 5 consecutive days by region can be found with aggregated information per day. IPMA forecasts roughly 41 regions, both onshore and offshore. Every twelve hours, the forecast information on the website is updated, for each region. In this work, relevance was given to the "id weather type", for which a number is attributed, corresponding to a weather description, which can be observed in Table 3.

Table 3 – IPMA API ID weather type [37], to allow the construction of a correspondence map. The bold values correspond to the ones used for the 50 % battery SOC target of strategy 3, SCM+RR+WF.

| Number | Correspondence | Number | Correspondence |
| --- | --- | --- | --- |
| --- | -99 | *14* | ***Intermittent heavy rain*** |
| 0 | No information | 15 | Drizzle |
| 1 | Clear sky | *16* | ***Mist*** |
| 2 | Partly cloudy | *17* | ***Fog*** |
| 3 | Sunny intervals | **18** | **Snow** |
| *4* | ***Cloudy*** | 19 | Thunderstorms |
| *5* | ***Cloudy (High cloud)*** | 20 | Showers and thunderstorms |
| 6 | Showers | 21 | Hail |
| 7 | Light showers | 22 | Frost |
| *8* | *Heavy showers* | *23* | *Rain and thunderstorms* |
| 9 | Rain | *24* | *Convective clouds* |
| 10 | Light rain | *25* | *Partly cloudy* |
| *11* | *Heavy rain* | *26* | *Fog* |
| 12 | Intermittent rain | *27* | *Cloudy* |
| 13 | Intermittent light rain | | |

With the help of this information, the battery SOC is prepared for the next day, as needed, through battery charging during the night hours (in general there is low energy consumption from domestic users during those hours). For that reason, this control type is optimal for the Portuguese bi-hourly and tri-hourly household tariffs, with lower electricity prices during the night [40].

*2.6.2. VRFB operation*

For continuous operation control and alarm detection, the battery terminals, cells of the stack, and the reference cell have installed electric sensors, which variables are real-time acquired. The variables are voltage and current, temperature (high accuracy probes), pressure, and mass flow, watt meters, and many possible alarms. The SOC of the battery is obtained in the course of the operation, through the real-time acquisition of the open-circuit voltage, as detailed in [36]. UÉvora's VRFB is generally operated at a depth of discharge of 85 %, in previous characterization tests it was possible to obtain a specific energy density of near $18.5 \pm 4.2$ Wh/Kg, a maximum useful capacity of $66.5 \pm 4$ kWh, a battery efficiency of $77.1 \pm 3.36$ %, and a response time of seconds [41]. Considering the small relative error of the model when calculating the VRFB key parameters, including the SOC, and the availability of power (charge/discharge), a battery operating range of 20% and 70% of SOC was selected (battery depth of discharge, DOD, of 50%), to avoid states of charge next to the extreme limits. The maximum power is constrained to the maximum inverter and battery limits and the real-time state of charge. The VRFB Power-SOC relation is considered for maximum charge and maximum discharge levels of power. Due to the Power-SOC characteristics of the BESS technologies the available power to charge/discharge is severely constrained near the upper and lower SOC limits. These available power technical restrictions hinder the full control of power ramps.

*2.6.3. Algorithm implementation*

The night charging only happens if the next day's forecast indicates a cloudy day, considering the map developed in Table 3. The cloudy day indicates a higher probability of the occurrence of PV power ramps. The flowchart of this algorithm part is presented in Figure 5. The ramp rate algorithm is activated when its real-time calculated result is equal to or larger than the 10 %/min of the PV nameplate capacity value. Every day at 1:30 am, the algorithm consults the IPMA API, to decide if it should act on the battery charging-only, to avoid the VRFB SOC being near to its lowest limit at the end of the day. If none of the conditions is satisfied - near lowest limit SOC, or "bad" weather day - the EMS continue its algorithm course, the self-consumption maximization.

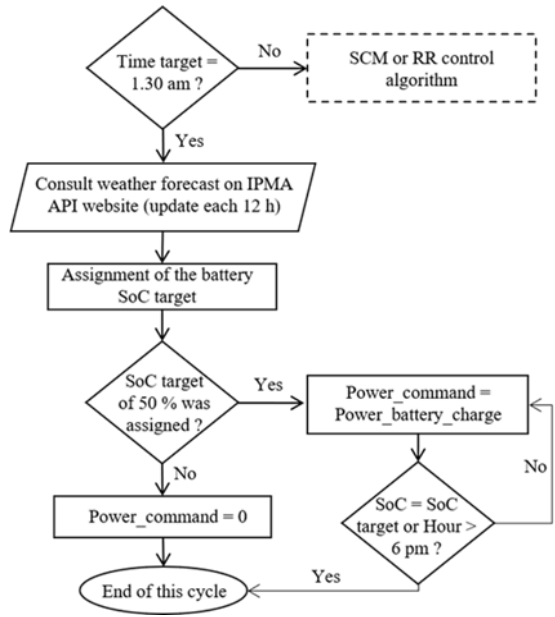

Figure 5 – Algorithm flowchart of the weather forecast with the IPMA API and SoC control. Where: Time_target – Initial hour for the beginning of the night charge (h); SoC target – Target to which the SoC should achieve, set as 50 %; Power_command – Power command value sent/received to/from the battery (W); and P_battery_charge – Constant charge power sent to charge the battery, in nightly hours (W).

Figure 6 presents the algorithm performing the ramp rate control when the defined maximum ramp value is violated. Given the inputs, at every 2 s the PV data is read lido and the ramp rate is calculated. If it violates the 10%/min of the PV nameplate capacity, the SOC is observed, and if possible, the battery compensates the PV deviation, with either a charge or a discharge (depending on the slope of the deviation). If the ramp rate violates the defined limit, the battery will only accommodate charges if its SOC is lower than the SOC lower limit, and discharges if its SOC is higher than the SOC higher limit. If the ramp rate is lower than the 10%/min of the PV nameplate capacity, the EMS becomes the self-consumption maximization (SCM). The SCM algorithm was presented and validated in the work of [11], wherefore will not be detailed in this work.

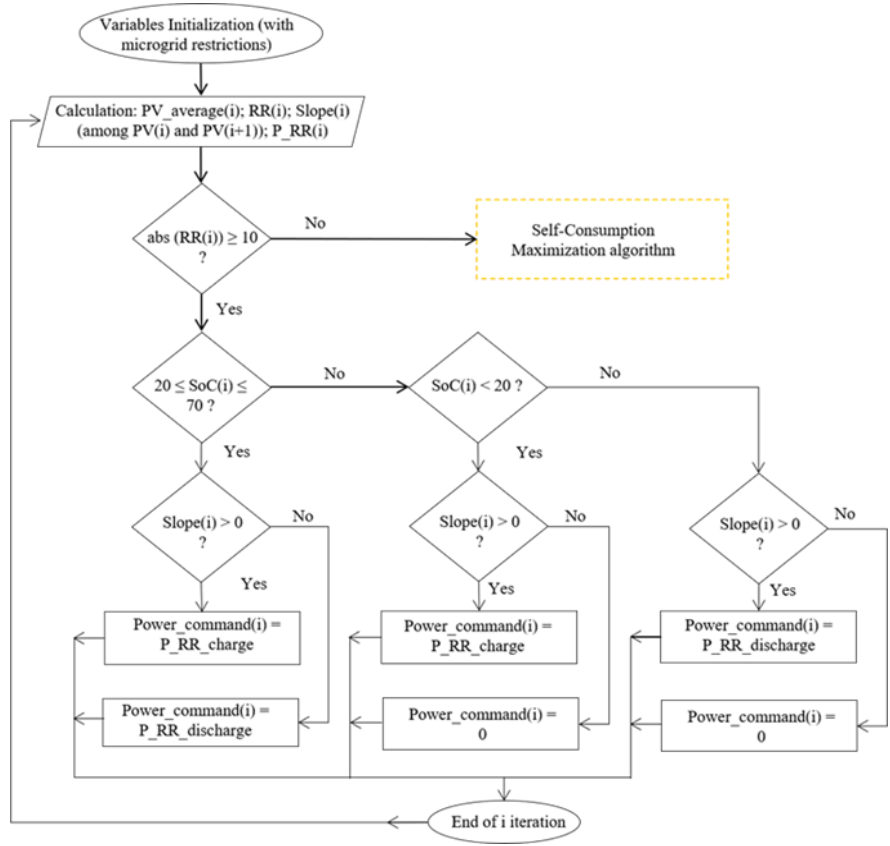

Figure 6 - Ramp-rate control algorithm flowchart, highlighting the battery power command power calculated at each iteration, with a cycle time frame of 2s. Where: i – Cycle iteration number; PV_average – PV values average of samples correspondent to 20 s (W); P_RR – Equivalent to the $P_{battery}$ of Equation (2) presented in Section 2.4 (P_RR_charge - $P_{battery}$ is a battery charge command and P_RR_discharge - $P_{battery}$ is a battery discharge command; and Slope – Slope of the two consecutive values of PV.

## 3. Results and Discussion

This section summarizes the main results and its related discussion obtained within this study, considering the consulted and described studies mentioned in Section 1 and the purpose of this investigation.

### 3.1. Weather Forecast VRFB charging

Power ramp rate control earns value if a state of charge control is implemented. Figure 7 presents the impact of having this battery charging for the SOC management in the current SOC of the battery over the course of the studied days where the control is applied. For the chosen week (1-7 January 2018), the IPMA API weather forecast was consulted for the next day forecasts, resulting in an active SOC control (night battery charge) on days 2, 3, 4 and 5. On days 1, 6, and 7, due to the weather forecast, the active SOC control was not activated.

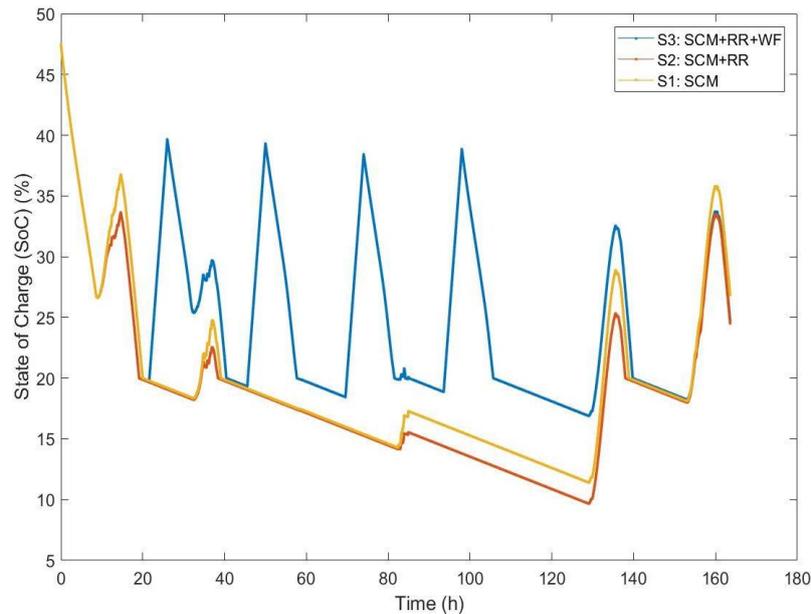

Figure 7 – Simulation of the battery SOC for the three strategies, over the same week PV-generation and load profiles.

If the VRFB SOC is near its extreme limits, the ramp rate control cannot be correctly applied, approaching the results of Strategy 2 to Strategy 1. The VRFB night charging based on the WF maximizes the amount of energy that flows to and from the VRFB (see TBU and FBU indicators, further discussed). Strategy 3, SCM+RR+WF, allows to deal with a bigger percentage of PV fluctuations, can successfully control all power ramps (CRR) with a ramp rate limit of 10%/min. On the other hand, Strategy 2 only control about 86% of the total power ramps occurring in the test period.

The WF control type relates the geographical location of the PV installation with the management of the battery. The developed approach allowed the battery SOC to be effectively controlled, allowing enough energy capacity for the next day's absorption or injection PV ramping needs. Improvements in the forecasts will impact the improvement of the EMS. The night battery charging power setpoint should be adapted for the considered technology type and energy capacity. The night charging is attractive for bi-hourly and tri hourly tariffs, given the cheapest price for off-peak energy. The SOC management could be achieved through other algorithm types, such as linear or dynamic ramp-rate limiter control, depending on the required speed and efficiency of the application. The WF VRFB charging method, based on locally projected forecasts proves to be a good method to use in Portugal, and certainly could be reproduced in other countries. A further sensitivity analysis of the SOC different controls should be addressed in future building sector studies.

### 3.2. Energy KPIs

The KPIs were calculated for each of the simulation environments, for the one week of January. For the reader to be engaged with the significance of the key performance indicators, a best-case scenario for the prosumer (self-consumption user/installation owner) point-of-view is presented. This reference value can help the reader to understand how close or distant a strategy is from its ideal (best-case) scenario. The best-case for one indicator could imply the worst case of another. The week chosen in this study can also influence some of the KPIs. The results can be observed in Table 4, presented below.

Table 4 – Results obtained of the KPIs and their respective best-case-scenario, for the simulation study, of the 1st week of January 2018.

| KPI | 1 SCM (%) | 2 SCM+RR (%) | 3 SCM+RR + WF (%) | Ideal best value (%) | Ideal best value direct meaning, from the point of view of the prosumer |
|---|---|---|---|---|---|
| SCR | 58.6 | 58.7 | 59.2 | 100 | The amount of PV energy produced meets with the PV energy consumed |
| SSR | 21.8 | 22.9 | 21.8 | 100 | The amount of PV energy consumed meets the energy consumed |
| GRF | 57.9 | 63.3 | 60.8 | 100 | Level of independence from the grid, from the user point-of-view (if equal to 100 %, there is no energy to the grid). |
| BCR | 38.7 | 37.4 | 40.9 | 50 (with 100% of battery efficiency) | Maximum possible energy used to charge the battery. Dependent on the battery energy capacity |
| EG | 98.8 | 95.2 | 94.9 | 100 (should be weighted) | Quantification of the amount of energy coming from the grid or going to the grid. Equal to 100 % means the energy sent to the grid is 0 %. |
| FGU | 57.2 | 60.2 | 57.7 | 0 | Equal to 100 % means that all the overall consumed energy comes from the grid. |
| TGU | 0.71 | 3.06 | 3.09 | 0 | Equal to 100 % means that the energy injected into the grid is equal to the load profile. |
| FBU | 13.2 | 10.8 | 22.2 | 100 | Equal to 100 % means that all the consumed energy comes from the battery. |
| TBU | 20.9 | 18.1 | 32.0 | 0 | Equal to 100 % means that all the energy sent to the battery (charge) is equal to the load energy needs. |
| CRR | 0.00 | 85.9 | 100 | 100 | Equal to 100 % means all ramps that violate the ramp rate reference are controlled. |

To improve the readability of the previous enunciated indicators, a graphical representation is presented in Figure 8 with the results of the resultant KPIs of the EMSs and the best-case scenario from the point-of-view of the prosumer.

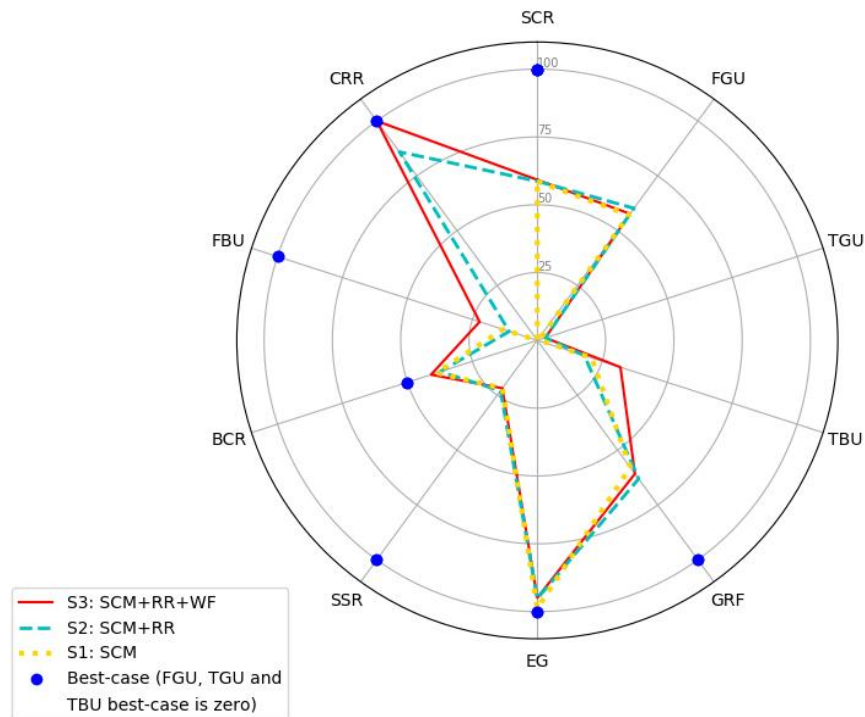

Figure 8 – Representation of the results of the KPIs for each of the simulated strategies, with the best-case scenario illustration inclusion.

Given this configuration context and the obtained best-case values, no strategies could achieve it, although some can get close. From the KPI results for all the strategies, the SCR parameter is consistent. Besides the PV-generated energy that is consumed directly by the installation, SCR is dependent on the availability of the battery to charge or discharge, and the main reason that the strategy 2 and 3 present a slightly higher value than the approach of strategy 1. Considering the understudy week, SCR is far from its theoretical maximum due to the PV generation of energy being greater than the global energy load profile and suffering a wide variation throughout the entire year (seasonality). The energy used for ramp control tasks (PV or battery) is lower when compared with the consumption needs. This means that this objective does not greatly penalize the SCR KPI.

The SSR indicator is curiously the same on strategy 1 and strategy 3, and higher than strategy 2. Similar behaviour to the previous indicator is observed regarding the best-case scenario. This indicates the ramp rate control does not affect the consumed PV energy, which was a priority to follow within the three strategies. Given the PV-generation and load profiles, the GRF is marginally higher for strategy 2 (SCM+RR). This KPI relates the energy extracted from the grid and the energy needed to supply the load. The night battery charging of strategy 3 could have led to the highest total energy exchanged to the grid, which did not happen. The strategy with the highest EG parameter in the overall grid use is Strategy 1 (SCM) mainly due to a lower weekly total of energy exchanged with the grid. The TGU indicator presents a growth from the simplest to the most complex strategy, meaning increased energy injected into the grid, although the grid injection of strategies 2 and 3 is smoothed since the ramp rate control is activated whenever a violation of the ramp rate limit occurs. TGU results offered a near value to its appointed best-case scenario. The energy extracted from the grid over the load profile is represented by FGU, which presents a distant value from the pointed best-case. Strategy 2 presents the highest value since it makes less use of the battery interaction with the loads (when devoted to the ramp rate control). BCR addresses the energy used to charge the battery over the total energy exchanged with the battery. It offers a clear representation of the energy fluxes in and out of the battery, and its pointed best-case is close to the obtained results.

This work focused to evaluate energetic KPIs, although the importance of economic indicators is recognized. Nevertheless, it was intended to minimize the cost of charging the VRFB (to control SOC when needed) overnight using the cheapest electricity tariffs (bi-hourly or tri-hourly tariffs).

*3.2.1. Type of battery impact on energy indicator – ageing and capacity fading*

The implementation of the WF VRFB charge for SOC ramp rate control causes an increase in the battery utilization rate, as expected, noticeable in the increase of the TBU and FBU indicators. In the case of the VRFB, this additional usage will not reduce its lifespan or increase its degradation, which could happen, for example, in lithium-ion battery (LIB) technology. The MA technique is a satisfactory method to approach the ramp rate calculation and was satisfactorily implemented in this VRFB. MA implies more cycling numbers than other ramp rate techniques, which is not an obstacle for this work since this VRFB presents a considered nominal energy capacity (60 kWh).

3.3. Combination of distinct controls aims

The PV self-consumption maximization energy management strategy could improve certain desired indicators, without compromising the main issues for which the strategy was built to solve. The analysis made concerns one week in the Portuguese winter season, a season usually characterized by several daily fluctuations in the PV power generation, and by the time of year with the lowest average daily global solar radiation. Figure 3 depicts the PV power generation over the period studied, with days characterized by high PV variability (except for day 6), posing a challenging scenario concerning power ramps. Combining distinct controls is to combine distinct aims, which should be carefully analysed to maintain the primary objective and not conduct to misleading results. In this work case, the combined strategy provides solves issues of grid quality, does not worse the PV self-consumption rate.

3.4. VRFB charge based on WF Validation Analysis

One important demand of the energy management strategies real-time application is the response of the system to the algorithms' commands, due to the importance of the simulation versus actuation. The online operation complicates the task, given the reliance on the data acquisition system and its real-time monitoring. Through the validation, the algorithm's performance at real-scale and real-time operation was possible to evaluate, accounting for technical constraints, data logging periods, the performance of the API of the WF, efficiencies of the equipment, and communication delays. A time frame of half a second and one second of cycling was tested, although the many variables controlled did not have time to execute all the commands, and by this means, a control cycle time of about 2 seconds was the minimum possible time frame to execute the algorithm properly. The comparison of the validation with the simulation enables the improvement of the technical variables considered in the model under study. Figure 9, below, depicts the VRFB SOC over the validation week, so the difference between the model and the operation can be briefly checked, and the validation of the algorithm using VRFB could be assessed.

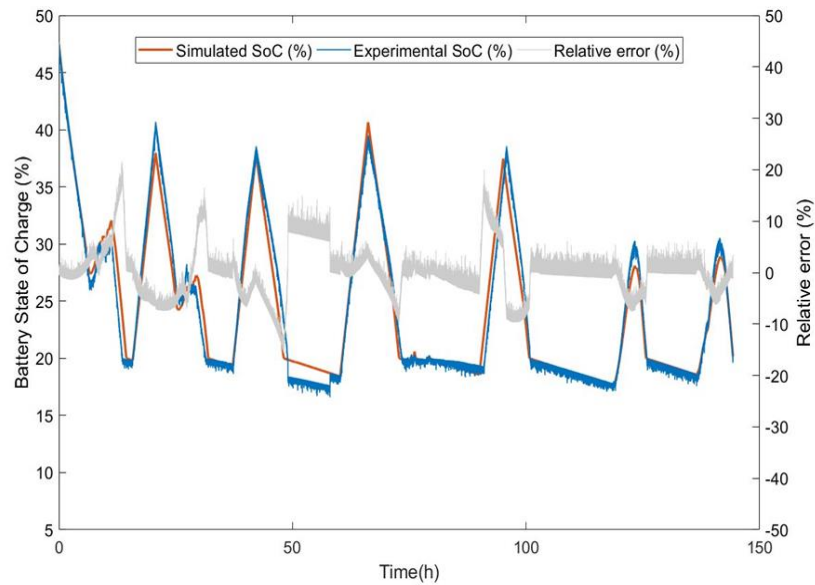

Figure 9 – SOC of the battery evolution over the period studied: SCM+RR+WF simulation vs. real-time implementation results, with a mean absolute error of 3.5%.

Considering the night battery charge in a mean value of 2700 W, and the PV-generation and load profiles, the battery SOC never exceeds 50 % for any strategy throughout the studied week. The difference between the simulation and the experimental output presents a low error (absolute average of 3.5 % over the period of study). Given the overall mentioned errors of real-time application of Strategy 3, which relies on the everyday WF, one can conclude that the EMS is possible to implement with the real-time control using the VRFB as BESS daily, and even with distinct control objectives.

3.5. Limitations and further work

Within the MA, the time frame has a great impact on results, as observed in Figure 4. The 20 s average time frame allowed a high number of maximum controlled ramps and does not present a high influence by the averaging of several different measurements of PV generation fluctuations. The WF VRFB charging power setpoint should be adapted for the technology type and energy capacity. To improve confidence in this ramp rate occurrence distribution, it is necessary to collect data from a wider period, in a similar way to a Typical Meteorological Year (TMY). Its application in the generation and distribution systems is an option, which needs to be assessed. Besides the technic and energy approach followed by this work, economic evaluation still needs to be addressed. Reduction of costs will be influenced by recent analysis cost, simulation, overall efficiency optimization, anticipated flaws detection, systems characterization. Improvements in the model simulation could benefit the accuracy of the results. Economic analysis and future impacts of application in the grid should be addressed in future work. Other promising SOC management controls, to attain the similarities in aim, could be associated with, for example, seasonality.

## 4. Conclusion

This study assesses a building scenario with a PV installation, a VRFB as the electricity storage unit, and a load profile over one week of wintertime, with data from January 2018. Based on the predicted growth of the PV penetration at building scenario for the years to come, three EMSs were simulated to obtain a result improvement of the main KPIs related to the self-consumption maximization and power ramp rate control.

For the year 2018, and based on local data, it was shown that about 5% of the PV power ramps that occurred were above the rate of 10%/min of the PV nameplate capacity understudy. PV fluctuations should be carefully addressed for the buildings sector, given the increase of solar PV installations in buildings. Strategy 1, SCM, performs a simple self-consumption maximization of the PV power generation; strategy 2, SCM+RR, additionally performs a ramp rate control, imposing a 10 %/min of the PV nameplate capacity RR limit, providing additional stability over the grid energy exchange. Strategy 3, SCM+RR+WF, added a 12-hour WF based on the IPMA API, to implement a SOC control able to prepare the battery to better deal with the next-day PV power ramps. The night VRFB charging based on the WF presented an approach to follow to condition the SOC of the battery at the end of the day.

Despite the challenging scenario of occurrence of power ramps in the selected week, the SCM+RR+WF strategy demonstrated to be able to control 100% of the ramps with rates above 10%/min, maintaining the PV SCR (61%), and being able to keep the GRF close to a value of 68%, and being implemented successfully, achieving the proposed objectives of this work. The development of multi-objective EMSs, often with competing goals such as the SCM+RR+WF, presents different system needs for power, energy, usage cycles, or response time.

## Acknowledgements

The authors would like to thank the support of this work, developed under the European POCITYF project, financed by 2020 Horizon under grant agreement no. 864400. This work was also supported by the PhD. Scholarship (author Ana Foles) of FCT – Fundação para a Ciência e Tecnologia –, Portugal, with the reference SFRH/BD/147087/2019.

## 5. References

[1]   Adrian Whiteman, S. Rueda, D. Akande, N. Elhassan, G. Escamilla, and I. Arkhipova, "Renewable Capacity Statistics 2020," 2020.

[2]   Arina Anisie *et al.*, "Innovation landscape for a renewable-powered future: Solutions to integrate variable renewables," 2019.

[3]   T. Ackermann, N. Martensen, T. Brown, P.-P. Schierhorn, F. G. Boshell, and M. Ayuso, "Scaling Up Variable Renewable Power: The Role of Grid Codes," p. 106, 2016.

[4]   V. Gevorgian, M. Baggu, and D. Ton, "Interconnection Requirements for Renewable Generation and Energy Storage in Island Systems: Puerto Rico Example: Preprint," no. May, 2017.

[5]   Q. Zheng, J. Li, X. Ai, J. Wen, and J. Fang, "Overview of grid codes for photovoltaic integration," *2017 IEEE Conf. Energy Internet Energy Syst. Integr. EI2 2017 - Proc.*, vol. 2018-Janua, pp. 1–6, 2017.

[6]   J. M. Alvarez, I. de la P. Laita, L. M. Palomo, E. L. Pigueiras, and M. G. Solano, "Grid integration of large-scale PV plants: dealing with power fluctuations," *Large Scale Grid*

*Integr. Renew. Energy Sources*, pp. 131–170, 2017.

[7] Circutor, "Circutor," *CVM-1D Series*. [Online]. Available: http://circutor.com/en/products/measurement-and-control/fixed-power-analyzers/power-analyzers/cvm-1d-series-detail. [Accessed: 14-Dec-2020].

[8] W. A. Omran, M. Kazerani, and M. M. A. Salama, "Investigation of Methods for Reduction of Power Fluctuations Generated From Large Grid-Connected Photovoltaic Systems," *IEEE Trans. Energy Convers.*, vol. 26, no. 1, pp. 318–327, 2011.

[9] I. de la Parra, J. Marcos, M. García, and L. Marroyo, "Control strategies to use the minimum energy storage requirement for PV power ramp-rate control," *Sol. Energy*, vol. 111, pp. 332–343, 2015.

[10] M. R. Islam, F. Rahman, and W. Xu, *Advances in Solar Photovoltaic Power Plants*. Berlin Heidelberg: Springer Nature, 2016.

[11] I. de la Parra, J. Marcos, M. García, and L. Marroyo, "Dealing with the implementation of ramp-rate control strategies - Challenges and solutions to enable PV plants with energy storage systems to operate correctly," *Sol. Energy*, vol. 169, no. March, pp. 242–248, 2018.

[12] V. Musolino, C. Rod, P. J. Alet, A. Hutter, and C. Ballif, "Improved ramp-rate and self consumption ratio in a renewable-energy-based DC micro-grid," in *2017 IEEE 2nd International Conference on Direct Current Microgrids, ICDCM 2017*, 2017, pp. 564–570.

[13] J. Martins, S. Spataru, D. Sera, D. I. Stroe, and A. Lashab, "Comparative study of ramp-rate control algorithms for PV with energy storage systems," *Energies*, vol. 12, no. 7, 2019.

[14] A. Ellis and D. Schoenwald, "PV Output Smoothing with Energy Storage," 2012.

[15] A. Allik, H. Lill, and A. Annuk, "Ramp rates of building-integrated renewable energy systems," *Int. J. Renew. Energy Res.*, vol. 9, no. 2, pp. 572–578, 2019.

[16] República Portuguesa, *Decreto-Lei n.º 162/2019 de 25 de outubro*. Portugal, 2019, pp. 45–62.

[17] A. Foles, L. Fialho, and M. Collares-Pereira, "Techno-economic evaluation of the Portuguese PV and energy storage residential applications," *Sustain. Energy Technol. Assessments*, vol. 39, no. March, p. 100686, 2020.

[18] J. Moshövel *et al.*, "Analysis of the maximal possible grid relief from PV-peak-power impacts by using storage systems for increased self-consumption," *Appl. Energy*, vol. 137, pp. 567–575, 2015.

[19] G. Lorenzi and C. A. S. Silva, "Comparing demand response and battery storage to optimize self-consumption in PV systems," *Appl. Energy*, vol. 180, pp. 524–535, 2016.

[20] J. Weniger, J. Bergner, and V. Quaschning, "Integration of PV power and load forecasts into the operation of residential PV battery systems," in *4th Solar Integration Workshop*, 2014, pp. 383–390.

[21] Y. Riffonneau, S. Bacha, F. Barruel, and S. Ploix, "Optimal power flow management for grid connected PV systems with batteries," in *IEEE Transactions on Sustainable Energy*, 2011, vol. 2, no. 3, pp. 309–320.

[22] R. Luthander, J. Widén, J. Munkhammar, and D. Lingfors, "Self-consumption enhancement and peak shaving of residential photovoltaics using storage and curtailment," *Energy*, vol. 112, pp. 221–231, 2016.

[23] G. L. Kyriakopoulos and G. Arabatzis, "Electrical energy storage systems in electricity generation : Energy policies , innovative technologies , and regulatory regimes," *Renew. Sustain. Energy Rev.*, vol. 56, pp. 1044–1067, 2016.

[24] R. Khalilpour and A. Vassallo, "Planning and operation scheduling of PV-battery systems: A novel methodology," *Renew. Sustain. Energy Rev.*, vol. 53, pp. 194–208, 2016.

[25] P. Alotto, M. Guarnieri, and F. Moro, "Redox flow batteries for the storage of renewable energy: A review," vol. 29, pp. 325–335, 2014.

[26] M. Guarnieri, P. Mattavelli, G. Petrone, and G. Spagnuolo, "Vanadium Redox Flow Batteries," *IEEE Ind. Electron. Mag.*, no. december, pp. 20–31, 2016.

[27] World Energy Council, "World Energy Resources 2016," London, 2016.

[28] K. Mongird, V. Viswanathan, J. Alam, C. Vartanian, V. Sprenkle, and R. Baxter, "2020 Grid Energy Storage Technology Cost and Performance Assessment," 2020.

[29] B. Turker *et al.*, "Utilizing a vanadium redox flow battery to avoid wind power deviation penalties in an electricity market," *Energy Convers. Manag.*, vol. 76, pp. 1150–1157, 2013.

[30] T. Sarkar, A. Bhattacharjee, H. Samanta, K. Bhattacharya, and H. Saha, "Optimal design and implementation of solar PV-wind-biogas-VRFB storage integrated smart hybrid microgrid for ensuring zero loss of power supply probability," *Energy Convers. Manag.*, vol. 191, no. January, pp. 102–118, 2019.

[31] A. Bhattacharjee, H. Samanta, N. Banerjee, and H. Saha, "Development and validation of a real time flow control integrated MPPT charger for solar PV applications of vanadium redox flow battery," *Energy Convers. Manag.*, vol. 171, no. June, pp. 1449–1462, 2018.

[32] A. H. Fathima and K. Palanismay, "Modeling and Operation of a Vanadium Redox Flow Battery for PV Applications," *Energy Procedia*, vol. 117, pp. 607–614, 2017.

[33] A. Foles, L. Fialho, M. Collares-Pereira, and P. Horta, "Vanadium Redox Flow Battery Modelling and PV Self-Consumption Management Strategy Optimization," in *EU PVSEC 2020 - 37th European Photovoltaic Solar Energy Conference and Exhibition*, 2020.

[34] EDP Distribuição, "Atualização dos perfis de consumo, de produção e de autoconsumo para o ano de 2018 Documento Metodológico," 2017.

[35] F. P. M. Kreuwel, W. H. Knap, L. R. Visser, W. G. J. H. M. van Sark, J. Vilà-Guerau de Arellano, and C. C. van Heerwaarden, "Analysis of high frequency photovoltaic solar energy fluctuations," *Sol. Energy*, vol. 206, no. May, pp. 381–389, 2020.

[36] L. Fialho, T. Fartaria, L. Narvarte, and M. C. Pereira, "Implementation and validation of a self-consumption maximization energy management strategy in a Vanadium Redox Flow BIPV demonstrator," *Energies*, vol. 9, no. 7, 2016.

[37] IPMA, "IPMA API." [Online]. Available: http://api.ipma.pt/open-data/forecast/meteorology/cities/daily/. [Accessed: 14-Dec-2020].

[38] European Centre for Medium-Range Weather Forecasts, "ECWMF." [Online]. Available: https://www.ecmwf.int/. [Accessed: 14-Dec-2020].

[39] National Centre for Meteorological Research, "AROME." [Online]. Available: https://www.umr-cnrm.fr/spip.php?article120&lang=en. [Accessed: 14-Dec-2020].

[40] EDP Comercial, "EDP Comercial," *Electricity Tariffs*. [Online]. Available: https://www.edp.pt/particulares/energia/tarifarios/. [Accessed: 14-Dec-2020].

[41] E. López, L. Fialho, L. V. Vásquez, A. Foles, J. S. Cuesta, and M. C. Pereira, "Testing and evaluation of batteries for commercial and residential applications in AGERAR project," in *Mission 10 000: BATTERIES*, 2019.